# The Japan earthquake of March 11[th], 2011 (Mw = 8.9R) as viewed in terms of local lithospheric oscillation due to M1 and K1 tidal components. A brief presentation.

Thanassoulas[1], C., Klentos[2], V., Verveniotis, G.[3], Zymaris, N.[4]

1. Retired from the Institute for Geology and Mineral Exploration (IGME), Geophysical Department, Athens, Greece.
   e-mail: thandin@otenet.gr - URL: www.earthquakeprediction.gr

2. Athens Water Supply & Sewerage Company (EYDAP),
   e-mail: klenvas@mycosmos.gr - URL: www.earthquakeprediction.gr

3. Ass. Director, Physics Teacher at 2[nd] Senior High School of Pyrgos, Greece.
   e-mail: gver36@otenet.gr - URL: www.earthquakeprediction.gr

4. Retired, Electronic Engineer.

**Abstract**

The time of occurrence of the large EQ that occurred recently in Japan (March 11[th], 2011, Mw = 8.9) is compared to the time of peak amplitude occurrence of the M1 and K1 tidal components. It is shown that the specific EQ occurred on the peak of the M1 tidal component, and deviates for only 45 minutes from the corresponding K1 tidal peak. Therefore, the specific seismic event complies quite well with the earlier proposed physical mechanism (lithospheric oscillation) that causes triggering of large EQs.

**Key words:** Japan, large earthquakes, M1 tidal wave, K1 tidal wave, lithospheric oscillations, tidal oscillations short-term earthquake prediction.

**1. Introduction.**

The aim of this very brief presentation is to show that the M1 and K1 tidal components play an important role concerning the time of occurrence of a large EQ. Actually, they provide the last decisive bit of stress load required in order to trigger a large EQ at an already critically stress charged seismogenic area. The physical mechanism that holds for the EQ triggering by the tidal waves has been presented in detail by Thanassoulas (2007) while specific examples have been presented from the Greek seismogenic area by Thanassoulas (2007), Thanassoulas and Klentos (2010), Thanassoulas et al. (2011) and from New Zealand (Thanassoulas et al. 2011).

In this brief presentation the time of occurrence of the very large EQ (Mw = 8.9) that occurred on 11[th] of March, 2011 in Japan will be compared to the local seismogenic area tidal conditions within a time window of some days before and after the EQ occurrence time. The tidal data have been determined by the Rudman et al. (1977) method.

**2. The large EQ (Mw = 8.9R) of Japan of the 11[th] of March, 2011.**

The location of the large EQ is shown in the following figure (1) as presented by the EMSC.

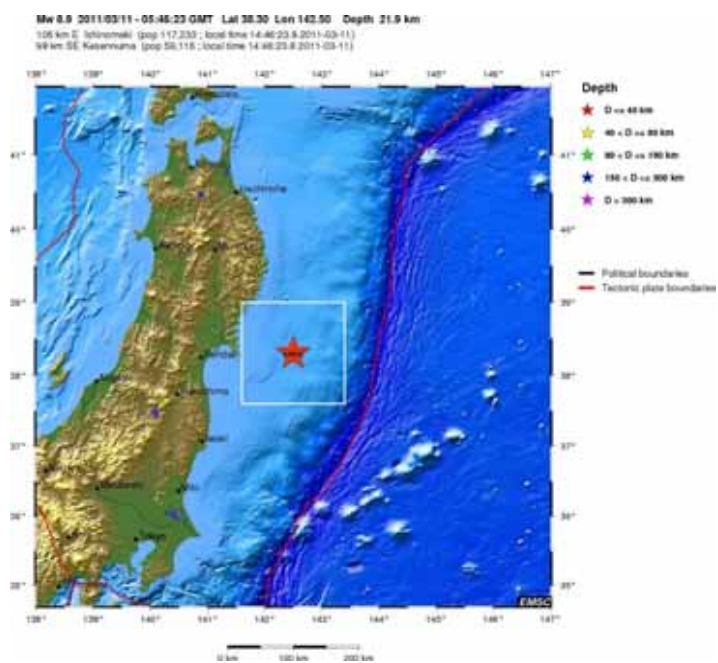

Fig. 1. Location (red star) of the large EQ (Mw = 8.9) of the 11[th] of March, 2011 in Japan.

**The corresponding M1 (T = 14 days) component tidal data.**

The M1 tidal oscillating component will be compared to the time of occurrence of the corresponding large EQ. The comparison is presented in the following figure (2).

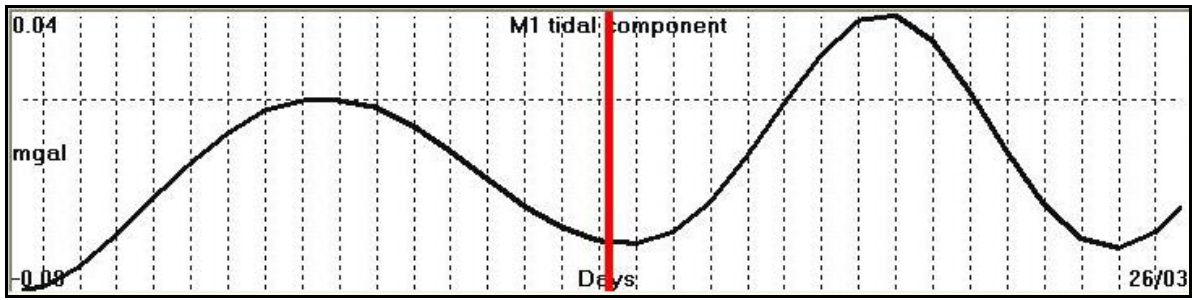

Fig. 2. Comparison of the M1 tidal oscillation (black line, the lithosphere is forced to oscillate in the same mode) with the time of occurrence (red bar) of the EQ of 11th of March, 2011 (Mw = 8.9). Vertical scale is in mgals.

In this case the EQ occurred, compared to the lithospheric tidal oscillation, exactly on the peak day of the M1 tidal wave

**The corresponding K1 (T = 24 hours) tidal data.**

Next, the lithospheric oscillating K1 component will be compared to the time of occurrence of the corresponding large EQ. The comparison is presented in the following figure (3).

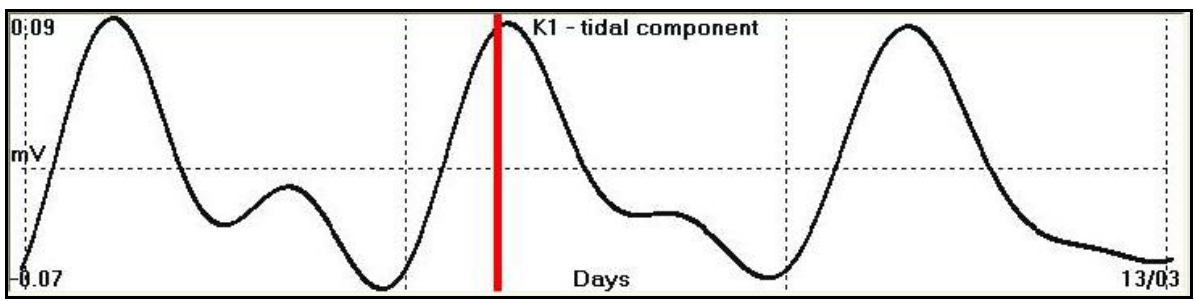

Fig. 3. Comparison of the K1 tidal oscillation (black line, the lithosphere is forced to oscillate in the same mode) with the time of occurrence (red bar) of the EQ of the 11th of March, 2011 (Mw = 8.9). Vertical scale is in mgals.

The specific EQ deviated for only forty five (45) minutes from the corresponding lithospheric oscillating tidal peak of the K1 component.

**3. Conclusions.**

The inspection of figures (2) and (3) shows that:

a) the recent large EQ (Mw = 8.9) of Japan occurred on the 11$^{th}$ of March of 2011 when the tidal component of M1 achieves an oscillation amplitude peak (minimum) and consequently, the corresponding seismogenic area reaches a maximum of stress load on this specific day. Therefore the deviation (dt) of the time of the EQ occurrence from the M1 tidal peak time is null.

**dt = 0 days in terms of M1 tidal component.**

b) the recent large EQ (Mw = 8.9) of Japan occurred on the 11$^{th}$ of March of 2011 and closely when the tidal component of K1 achieves a daily oscillation amplitude peak (maximum) and consequently the corresponding seismogenic area reaches a same day short-term maximum of stress load. The deviation (dt) of the time of the EQ occurrence from the tidal peak time is determined from figure (3) as dt = 45 minutes from the K1 tidal peak time.

**dt = 45 minutes in terms of K1 tidal component.**

Consequently, the large EQ of Japan that occurred on 11$^{th}$ of March of 2011 with a magnitude of Mw = 8.9 complied quite well in terms of its time of occurrence and the times of the peaks of the M1 and K1 tidal oscillating components with the physical mechanism earlier presented by Thanassoulas (2007).
A detailed description of the postulated physical mechanism and more representative examples can be found in **www.earthquakeprediction.gr**

**4. References.**

**EMSC: European – Mediterranean Seismological Centre, http://www.emsc-csem.org/#2**

**http://www.earthquakeprediction.gr**

**Rudman, J. A., Ziegler, R., and Blakely, R., 1977. Fortran Program for Generation of Earth Tide Gravity Values, Dept. Nat. Res. State of Indiana, Geological Survey.**

**Thanassoulas, C., 2007. Short-term Earthquake Prediction, H. Dounias & Co, Athens, Greece. ISBN No: 978-960-930268-5.**

**Thanassoulas, C., Klentos, V. 2010. How "Short" a "Short-term earthquake prediction" can be? A review of the case of Skyros Island, Greece, EQ (26/7/2001, Ms = 6.1 R).  arXiv:1002.2162 v1 [physics.geo-ph].**

**Thanassoulas, C., Klentos, V., Verveniotis, G., Zymaris, N. 2011. A short note on the two most recent large EQs of New Zealand (Mw = 7.0 on September 3rd 2010 and Mw = 6.1 on February 21st, 2011). A typical example of tidally triggered large EQs by the M1 tidal component.  arXiv:1102.4778 v3 [physics.geo-ph].**